\documentclass[11pt,preprintnumbers,aps,amssymb,nofootinbib,amsmath,superscriptaddress]
{revtex4}
\usepackage{epsfig,epsf}
\usepackage{bm} 
\usepackage{slashed}
\usepackage{color} 
\newcommand{\beq}{\begin{equation}}
\newcommand{\beql}[1]{\begin{equation}\label{#1}}
\newcommand{\eeq}{\end{equation}}
\def\bal#1\gal{\begin{align}#1\end{align}}
\newcommand{\ball}[1]{\bal\label{#1}}
\newcommand{\eq}[1]{(\ref{#1})}
\newcommand{\fig}[1]{Fig.~\ref{#1}}
\renewcommand{\sec}[1]{Sec.~\ref{#1}}
\newcounter{topiccounter}
\renewcommand{\b}[1]{{\bm #1}} 

\newcommand{\aver}[1]{\left\langle #1 \right\rangle}

\begin{document}

\title{Taming instability of magnetic field in chiral medium}
\author{Kirill Tuchin}

\affiliation{Department of Physics and Astronomy, Iowa State University, Ames, Iowa, 50011, USA}

\date{\today}


\begin{abstract}
 
Magnetic field is unstable in a medium with time-independent chiral conductivity.  Owing to the chiral anomaly, the electromagnetic field and the medium exchange helicity which results in time-evolution of the chiral conductivity. Using the fastest growing momentum and helicity state of the vector potential as an ansatz, the time-evolution of  the chiral conductivity and magnetic field is solved analytically. The solution for the hot and cold equations of state shows that the magnetic field does not develop an instability due to helicity conservation. Moreover, as a function of time, it develops a peak only if a significant part of the \emph{initial} helicity is stored in the medium. The initial helicity determines the height and position of the peak. 

\end{abstract}

\maketitle

\section{Introduction:  instability of magnetic field}\label{sec:i}

A medium with chiral anomaly responds to magnetic field by generation of anomalous electric current flowing in the magnetic field direction. The strength of this response is determined by the chiral conductivity $\sigma_\chi$. The induced current in turn produces the magnetic field in medium. This process obeys the Maxwell equations that yield an equation for the magnetic field
in a chiral medium  
\ball{i11}
-\nabla^2 \b B= -\partial_t^2\b B+\sigma_\chi \b \nabla\times \b B\,.
\gal
 Assuming that $\sigma_\chi$ is time-independent constant, a solution to \eq{i11} can be written as a superposition of the Chandrasekhar-Kendall states \cite{CK,Hirono:2015rla}. In particular, for a circularly polarized plane wave  with momentum $\b k$ 
\ball{i13}
\b B_{\b k\lambda}\sim e^{i\b k\cdot \b r }\exp\left\{-it\lambda_1\sqrt{k(k+\lambda\sigma_\chi)}\right\}\,,
\gal
where $\lambda_{1,2}= \pm 1$ (see e.g.\ \cite{Tuchin:2014iua,Manuel:2015zpa}). The state with $\lambda_1=1$ and $\lambda=-1$ grows exponentially with time at $k<\sigma_\chi$  indicating that the magnetic field is unstable in the chiral medium. This effect and its consequences have been discussed  in various contexts  \cite{Joyce:1997uy,Boyarsky:2011uy,Hirono:2015rla,Xia:2016any,Kaplan:2016drz,Kharzeev:2013ffa,Khaidukov:2013sja,Avdoshkin:2014gpa,Akamatsu:2013pjd,Kirilin:2013fqa,Tuchin:2014iua,Dvornikov:2014uza,Buividovich:2015jfa,Manuel:2015zpa,Sigl:2015xva,Kirilin:2017tdh}. 

The chiral conductivity is actually a function of time since energy and helicity are exchanged between the chiral medium and the magnetic field due to the chiral anomaly. It has been recently argued in \cite{Kaplan:2016drz} that the instability of the magnetic field is curtailed by the total energy and helicity conservation. The goal of this paper is to investigate this mechanism in more detail. To this end,  the vector potential is expanded into a complete set of the helicity states. The circularly polarized plane waves are used in this paper, but the same result can be obtained with any other Chandrasekhar-Kendall states as well.  The vector potential is then approximated by a single fastest growing (as a function of time) state.   Momentum $k_0$ of the fastest growing state  is proportional to the chiral conductivity and hence is a function of time. This is different from the monochromatic approximation discussed in \cite{Boyarsky:2011uy,Manuel:2015zpa}, where the momentum of a monochromatic state is constant. In the present model, the rate of the state growth is determined by the anomaly equation. 

In order to solve the Maxwell and the chiral anomaly equations one needs to know the equation of state of the medium that relates the chiral charge density to the axial chemical potential. We consider two equations of state \eq{g17} and \eq{n1} corresponding to  hot and cold media. In each case an analytical solution is derived for the chiral conductivity and magnetic field. The details of this derivation are discussed in the subsequent sections. The results are shown in \fig{fig:sigma-hot}--\fig{fig:b-cold-2}. One can see that the vector potential increases rapidly with time. However the magnetic field develops a maximum only if at the initial time a large enough fraction of the total helicity is stored in the medium rather than in the field. Eventually, the magnetic field vanishes at later times.

\section{Maxwell equations}\label{sec:a}

Electrodynamics coupled to the topological charge carried by the gluon field is governed by the Maxwell-Chern-Simons equations \cite{Wilczek:1987mv,Carroll:1989vb, Sikivie:1984yz,Kharzeev:2009fn} 
\bal
&\b \nabla\cdot \b B=0\,, \label{a15}\\
& \b \nabla\cdot \b E= \rho- c_A\, \b\nabla\theta\cdot \b B\,,  \label{a16}\\
& \b \nabla \times \b E= -\partial_t \b B\,,\label{a17}\\
& \b \nabla \times \b B= \partial_t \b E+ \b j + c_A(\partial_t\theta\, \b B+ \b\nabla \theta\times \b E)\,,\label{a18}
\gal
where $\theta$ is a pseudo-scalar field and  $c_A=N_c\sum_fq_f^2e^2/2\pi^2$, $N_c$ is the number of colors, $f$ is flavor index and $q_f$ is a quark electric charge in units of $e$. In a chiral medium, the time-derivative of $\theta$ can be identified with the axial chemical potential $\dot \theta=\mu_5$. We are going to consider an idealized case of a homogeneous medium $\b\nabla \theta=0$ with vanishing charge density  $\rho=0$. In this case the anomalous current is given by 
\ball{a20}
\b j_A= \sigma_\chi\b B\,,
\gal 
where the chiral conductivity defined as \cite{Fukushima:2008xe,Kharzeev:2009fn}
\ball{a22}
\sigma_\chi = c_A\mu_5
\gal
is a function of only time. In the radiation gauge $\b \nabla \cdot \b A=0$, $A^0=0$ Eq.~\eq{a18} can be written as an equation for the vector potential 
\ball{a24}
-\nabla^2 \b A= -\partial_t^2\b A+\b j+\sigma_\chi(t) \b \nabla\times \b A\,.
\gal

\section{The fastest growing state}\label{sec:b}

We proceed by expanding the vector potential into eigenstates of the curl operator $\b W_{\b k\lambda}(\b x)$, known also as the Chandrasekhar-Kendall (CK) states \cite{CK}.  A particular form of these functions is not important, but its easier to deal with them in Cartesian coordinates where they are represented by the circularly polarized plane waves
\ball{b11} 
\b W_{\b k\lambda}(\b x)= \frac{\b\epsilon^\lambda}{\sqrt{2kV}}e^{i\b k\cdot \b x}\,,
\gal
where $V$ is volume and $\lambda= \pm 1$ corresponds to the right and left polarizations. Functions \eq{b11} satisfy the eigenvalue equation 
\ball{b11.1}
\b\nabla\times \b W_{\b k\lambda}(\b x)= \lambda k \b W_{\b k\lambda}(\b x)
\gal
and the normalization condition 
\ball{b12}
\int \b W^*_{\b k\lambda}(\b x)\cdot  \b W_{\b k'\lambda'}(\b x)d^3x= \frac{1}{2k}\delta_{\lambda\lambda'}\delta_{\b k, \b k'}\,.
\gal
Expansion of the vector potential into a complete set of the CK states reads
\ball{b14}
\b A = \sum_{\b k, \lambda}\left[a_{\b k\lambda}(t)\b W_{\b k'\lambda'}(\b x)+ a^*_{\b k'\lambda'}(t)\b W^*_{\b k'\lambda'}(\b x)\right]\,.
\gal
The corresponding electric and magnetic fields are given by
\bal
&\b E= -\partial_t \b A = \sum_{\b k, \lambda}\left[-\dot a_{\b k\lambda}(t)\b W_{\b k'\lambda'}(\b x)- \dot a^*_{\b k\lambda}(t)\b W^*_{\b k\lambda}(\b x)\right]\,, \label{b16}\\
&\b B= \b\nabla\times  \b A = \sum_{\b k, \lambda}\left[\lambda k\, a_{\b k\lambda}(t)\b W_{\b k\lambda}(\b x)+\lambda k a^*_{\b k\lambda}(t)\b W^*_{\b k\lambda}(\b x)\right]\,. \label{b17}
\gal
Substituting \eq{b14} into \eq{a24} and using the Ohm's law $\b j = \sigma \b E$  one gets an equation 
\ball{b20}
k^2a_{\b k \lambda}= -\ddot a_{\b k \lambda}- \sigma \dot a_{\b k \lambda}+\sigma_\chi(t) \lambda k\, a_{\b k \lambda}\,.
\gal
It can be solved  in the adiabatic approximation, which is adequate for analysis of the unstable states \footnote{The self-consistency of this approximation is confirmed in Appendix,}.  Namely, we are seeking a solution in the form 
\ball{b22}
a_{\b k \lambda}= e^{-i\int_0^t\omega_{k\lambda}(t')dt'}
\gal
and assume that $\omega_{k\lambda}(t)$ is a slow varying function, which allows one to neglect terms proportional to $\dot \omega_{k\lambda}$. This yields
\ball{b24}
\omega_{k\lambda}(t)= \left[ -\frac{i\sigma}{2} + \lambda_1\sqrt{k^2-\sigma_\chi \lambda k-\sigma^2/4}\right]\,,
\gal
with $\lambda_1=\pm 1$. States with $\lambda_1=1$  and $k$ such that the expression under the square root is negative are unstable.  A more detailed  analyzes can be found in \cite{Tuchin:2014iua}.  We are going to concentrate on the fastest growing state, whose momentum $k_0$ corresponds to the maximum of the  function $-(k^2-\sigma_\chi \lambda k-\sigma^2/4)$. Namely, 
\ball{b25}
k_0= \frac{\sigma_\chi\lambda}{2}\,.
\gal
Clearly, $\sigma_\chi \lambda$ is positive in an unstable state. We assume that  $\sigma_\chi>0$ and $\lambda=1$.\footnote{If during the evolution $\sigma_\chi$ changes sign, then $\lambda=1$ state stops growing while $\lambda=-1$ becomes the fastest growing state.\label{fnt}} At $k=k_0$ Eq.~\eq{b24} becomes
\ball{b26}
\omega_0(t)= -\frac{i\sigma}{2}+\frac{i}{2}\sqrt{\sigma^2+\sigma_\chi^2(t)}\,.
\gal
Thus, the fastest growing state is 
\ball{b28}
a_0(t)= e^{\gamma(t)/2}\,, 
\gal
with
\ball{b29}
\gamma(t)=\int_0^t \left[ \sqrt{\sigma^2+\sigma_\chi^2(t')}-\sigma\right]dt' \,.
\gal
The model employed in the ensuing sections of this paper, consists in approximating the vector potential by the fastest growing mode given by \eq{b28},\eq{b29}. The corresponding vector potential is
\ball{b30}
\b A(\b r, t) \approx a_0(t)\b W_{\b k_0+}(\b r)+ \text{c.c.}
\gal
To verify that the ansatz Eq.~\eq{b30} is indeed a solution to Eq.~\eq{a24} one has to keep in mind that when taking the time-derivative of $\b A$, function $\b W_{\b k_0+}$ is treated as time-independent (even though $k_0$ depends on time), because its time-derivative is proportional to $\dot \omega_0$, which is neglected in the adiabatic approximation.

\section{Energy and helicity}\label{sec:d}

Now let us verify whether the energy conservation restricts the functional form of $\sigma_\chi(t)$.  Energy stored in the electromagnetic field is 
\ball{d3}
\mathcal{E}_\text{em}= \frac{1}{2} \int(\b E^2+\b B^2)d^3x= \sum_{\b k,\lambda}\frac{1}{2k}\left( |\dot a_{\b k\lambda}|^2+|a_{\b k\lambda}|^2k^2\right)\,,
\gal
where \eq{b16},\eq{b17},\eq{b12} were used. Similarly, magnetic helicity is given by \footnote{The gauge-dependence of magnetic helicity does not affect its time-evolution, see e.g.\ \cite{Biskamp}.}
\ball{d5}
\mathcal{H}_\text{em}= \int \b A\cdot \b B\, d^3x = \sum_{\b k,\lambda}\lambda |a_{\b k\lambda}|^2\,,
\gal
while the energy loss due to Ohm's currents reads
\ball{d7}
Q=\int \b j\cdot \b E \, d^3 x=\sigma \int \b E^2\, d^3x= \sigma \sum_{\b k,\lambda}\frac{1}{k} |\dot a_{\b k\lambda}|^2\,.
\gal

The energy balance equation  follows from Maxwell equations. Subtracting the scalar product of \eq{a17} with $\b B$ from the scalar product of \eq{a18} with $\b E$ and integrating over volume yields
\ball{d11}
\frac{1}{2}\partial_t \int(\b E^2+\b B^2)d^3x + \int \b j \cdot \b E\, d^3x+ \sigma_\chi (t) \int \b B\cdot \b E\,d^3x= 0\,,
\gal
where we neglected the surface contribution.  Noting that up to a surface term 
\ball{d13}
\partial_t \mathcal{H}_\text{em} = -2\int \b E\cdot\b B\, d^3x\,,
\gal
we can write \eq{d11} as 
\ball{d15}
\partial_t\mathcal{E}_\text{em}+ Q-\frac{1}{2} \sigma_\chi(t)\,\partial_t\mathcal{H}_\text{em} =0\,.
\gal

In the framework of our model, i.e.\ using the fastest growing state \eq{b30} as an ansatz for the vector potential, one obtains 
\bal
\mathcal{E}_\text{em}&= \frac{1}{2\sigma_\chi}\left( \sigma_\chi^2+\sigma^2-\sigma\sqrt{\sigma^2+\sigma_\chi^2}\right)e^{\gamma}\,,\label{d17}\\
\mathcal{H}_\text{em}&= e^{\gamma}\,,\label{d18}\\
Q &= \frac{\sigma}{2\sigma_\chi}\left( \sqrt{\sigma^2+\sigma_\chi^2}-\sigma\right)^2e^{\gamma}\,.\label{d19}
\gal
Noting that in the adiabatic approximation $\dot{\mathcal{E}}_\text{em}\approx \dot \gamma\mathcal{E}_\text{em}$ it is  straightforward to verify that Eqs.~\eq{d17}--\eq{d19} satisfy the energy balance equation \eq{d15}.

We emphasize that Eq.~\eq{d15} is satisfied for any function $\sigma_\chi(t)$. Thus, energy conservation does not tame the instability of the magnetic field. In particular, for a constant chiral conductivity, Eqs.~\eq{d17},\eq{d18},\eq{d19} diverge exponentially with time. This does not contradict the conclusions of Ref.~\cite{Kaplan:2016drz}, because it explicitly used the chiral anomaly equation.

\section{Chiral anomaly}\label{sec:g}

Time-dependence of $\sigma_\chi$ is determined by the chiral anomaly equation and the equation of state that connects the average  chiral charge density $\aver{n_A}$ to the axial chemical potential $\mu_5$. The chiral anomaly equation reads \cite{Adler:1969gk,Bell:1969ts}
\ball{g3}
\partial_\mu j_A^\mu =  c_A \b E\cdot \b B\,.
\gal
In  a homogeneous medium $\b \nabla \cdot \b j_A= \sigma_\chi \b \nabla\cdot \b B=0$, so that \eq{g3} reduces to an equation for the time component of the anomalous current
\ball{g5}
 \dot n_A=c_A \b E\cdot \b B\,.
\gal
Averaging over volume and using \eq{d13} gives\footnote{Our notations generally follow \cite{Hirono:2015rla}.} 
\ball{g9}
\partial_t\aver{n_A}=\frac{c_A}{V}\int \b E\cdot \b B\, d^3 x= -\frac{c_A}{2V}\partial_t\mathcal{H}_\text{em}\,.
\gal  
Integrating, one obtains the total helicity conservation condition
\ball{g11}
\frac{2V}{c_A}\aver{n_A}+\mathcal{H}_\text{em}= \mathcal{H}_\text{tot}\,,
\gal
where $ \mathcal{H}_\text{tot}$ is a constant.

Magnetic helicity $\mathcal{H}_\text{em}$ explicitly  depends on $\sigma_\chi(t)$, rather than on $\aver{n_A}$. Therefore, in order to solve \eq{g11} one needs an equation of state of the chiral medium that connects  the chiral charge density and the axial potential. It can be  computed from the grand canonical potential $\Omega$ as \cite{Fukushima:2008xe}
\ball{g13}
\aver{n_A}= -\frac{1}{V}\frac{\partial \Omega}{\partial \mu_5}\,.
\gal
Clearly, the equation of state depends on medium properties and, in general, is complicated even in the non-interacting approximation. It simplifies in two important limits: when temperatures $T$ and quark chemical potentials $\mu$ are much higher or  much lower than the axial chemical potential $\mu_5$. We refer to these limits as the hot and cold medium respectively and consider separately in the following two sections. 

\section{Magnetic field evolution in hot medium}\label{sec:k}

In a hot medium with $\mu,T\gg \mu_5$, the equation of state is linear \cite{Fukushima:2008xe}
\ball{g17}
\aver{n_A}= \chi\, \mu_5\,,
\gal
where the susceptibility $\chi$ depends on $\mu$ and $T$, but not on time.  It follows then from \eq{a22}  that 
\ball{g19}
\sigma_\chi(t)= \frac{c_A}{\chi}\aver{n_A(t)}\,,
\gal
The helicity conservation \eq{g11} now reads
\ball{g21}
\frac{\sigma_\chi(t)}{\alpha}= 1-\frac{\mathcal{H}_\text{em}(t)}{\mathcal{H}_\text{tot}}\,,
\gal 
 where $\alpha= \mathcal{H}_\text{tot}c_A^2/(2V\chi)$ is a characteristic energy scale. 
 
The vector potential in \eq{b30} is normalized such that at the initial time the magnetic helicity  equals unity $\mathcal{H}_\text{em}(0)=1$. Denoting the initial value of the chiral conductivity by $\sigma_\chi(0)= \sigma_0$ one can  infer from \eq{g21} that 
 \ball{g23}
 \sigma_0= \alpha(1-\mathcal{H}_\text{tot}^{-1})<\alpha\,.
 \gal
The ratio $\sigma'_0= \sigma_0/\alpha$ determines the fraction of the total \emph{initial} helicity stored in the medium.  
 If at  $t=0$ $\mathcal{H}_\text{tot}\gg 1$, then most of the initial helicity is stored in the medium, implying $\sigma_0'\lesssim 1$. Note that $\sigma_0'$ never equals 1 for a finite positive total helicity.\footnote{In the case of negative helicity, all terms in \eq{g11} would change sign, see footnote \ref{fnt}.} In the opposite case, all helicity is initially magnetic  $\mathcal{H}_\text{tot}= \mathcal{H}_\text{em}(0)=1$ implying $\sigma_0'=0$. 
 
 Substituting \eq{d18} into \eq{g21}, using the definition of $\gamma$ from \eq{b29} and taking the time-derivative, one derives an equation for $\sigma_\chi$
\ball{g25} 
\dot \sigma_\chi= -\left(\sqrt{\sigma^2+\sigma^2_\chi}-\sigma\right) (\alpha-\sigma_\chi)\,. 
\gal
It is convenient to use a set of dimensionless quantities 
\ball{g26}
\sigma'_\chi=\sigma_\chi/\alpha\,,\quad  \sigma'=\sigma/\alpha\,,\quad  \tau = \alpha t\,,
\gal
in terms of which Eq.~\eq{g25} is cast into the  form 
\ball{g27} 
\dot \sigma'_\chi= -\left(\sqrt{\sigma'^2+\sigma'^2_\chi}- \sigma'\right) (1-\sigma'_\chi)\,. 
\gal
In view of \eq{g23}, the right-hand-side of \eq{g27} is always negative. Perforce, $\sigma_\chi$ is a monotonically decreasing function of time implying that helicity always flows from the medium to the field until all of it is stored in the field. This is in contrast to \cite{Hirono:2015rla,Manuel:2015zpa} where the helicity can flow in both directions. How long it takes to transfer the helicity to the field depends on an equation of state as discussed in this and the following sections. 

\begin{figure}[ht]
\begin{tabular}{lcr}
      \includegraphics[height=4.5cm]{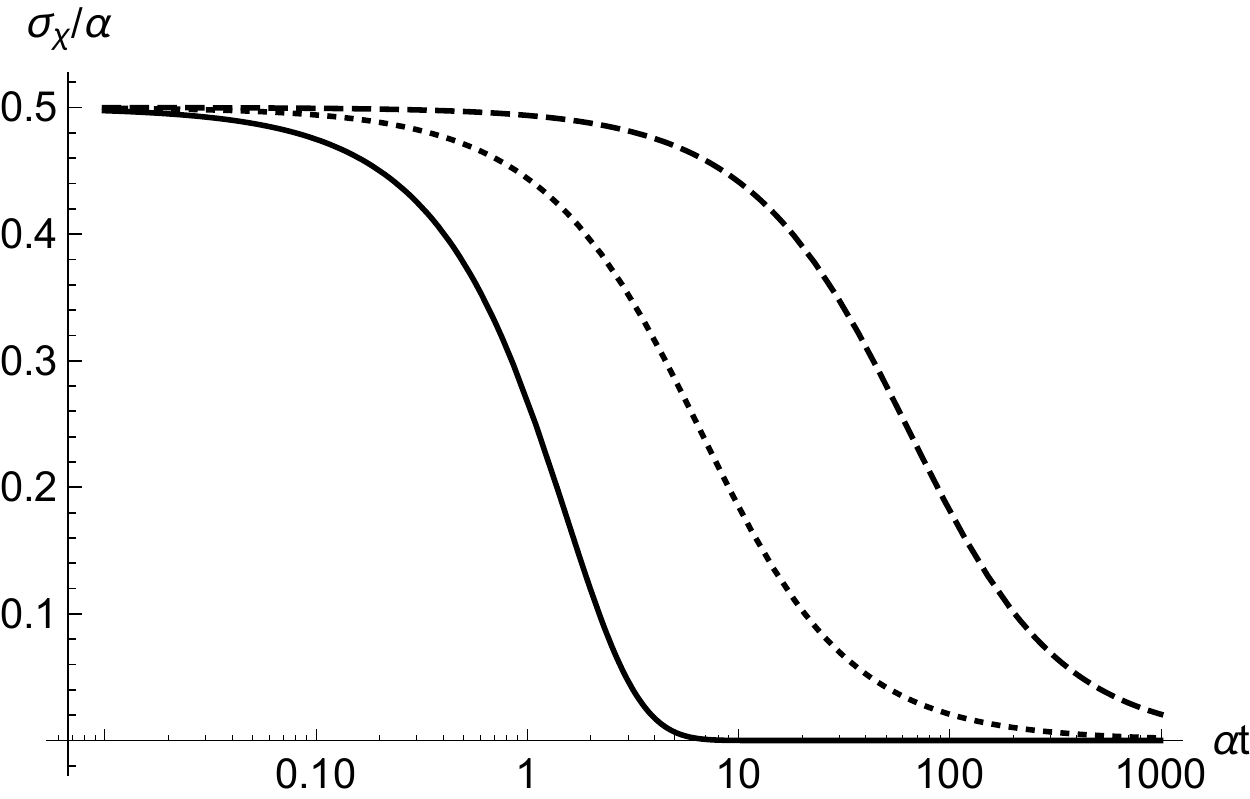}& &
      \includegraphics[height=4.5cm]{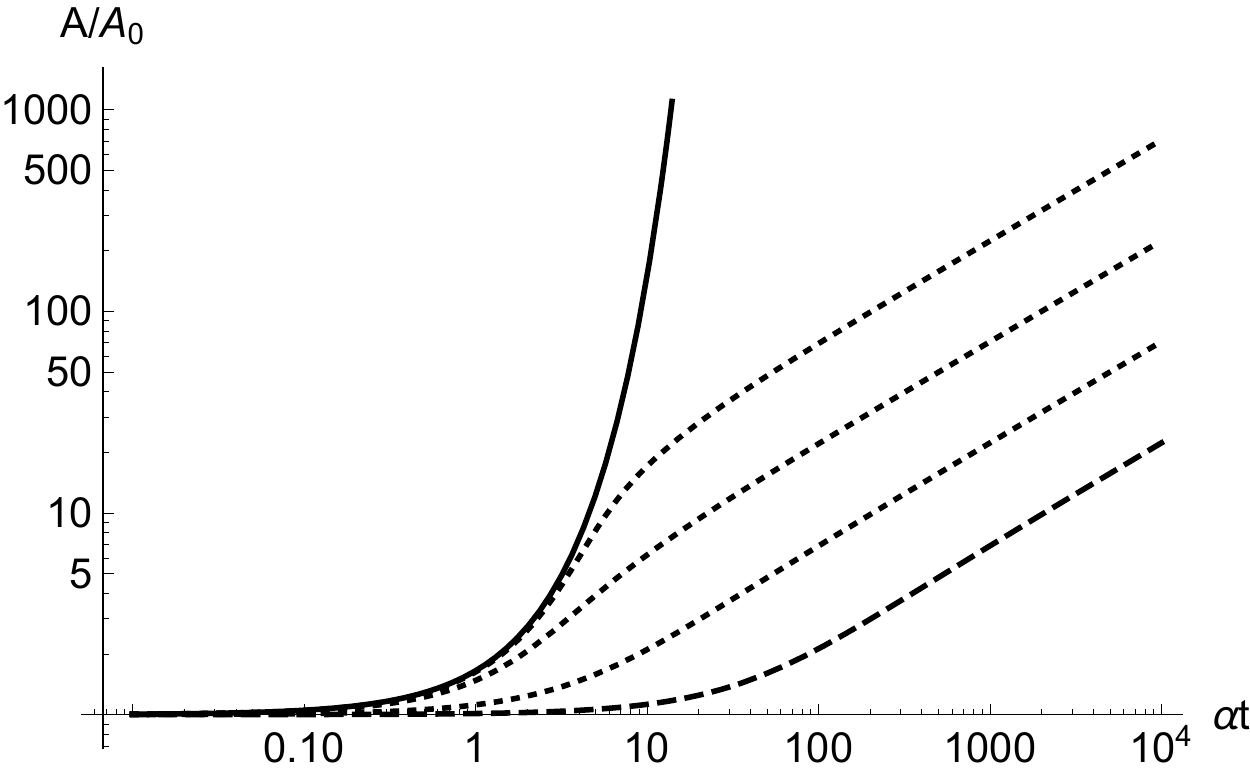}
      \end{tabular}
  \caption{Time dependence of the the chiral conductivity (left panel) and the  vector potential (right panel) in hot medium with the initial condition $\sigma'_\chi(0)\equiv \sigma'_0=0.5$ (50\% of the initial helicity is in the field). Lines on the left panel represent a numerical solution of \eq{g27} for   $\sigma'=0$ (solid line), $1$ (dotted line), $10$ (dashed line). Lines on the right panel represent \eq{k5} for $ \sigma'=0$ (solid line),   $0.01$, $0.1$, $1$ (dotted lines left-to-right), $10$ (dashed line).}
\label{fig:sigma-hot}
\end{figure}

Eq.~\eq{g27} can be analytically integrated and yields a  transcendental equation for $\sigma'_\chi(\tau)$. It is not very illuminating though, so instead we directly plot its solution in left panel of \fig{fig:sigma-hot}.

Once the chiral conductivity is determined from \eq{g27}, the vector potential can be computed using \eq{b30} and \eq{b11}. Its time-dependence is given by
\ball{g31}
A\propto \frac{1}{\sqrt{\sigma_\chi(t)}}a_0(t)\,.
\gal
Since $\gamma$ is an increasing function of time, $a_0$ and $A$ are also increasing functions of time, which is important for the model self-consistency. Time-dependence of $A$ is exhibited in the right panel of \fig{fig:sigma-hot}. The magnetic  helicity $\mathcal{H}_\text{em}=a_0^2$ grows from its initial value $\mathcal{H}_\text{em}(0)=1$ to the final value of $\mathcal{H}_\text{em}(\infty)=\mathcal{H}_\text{tot}$. 

\begin{figure}[ht]
\begin{tabular}{lcr}
      \includegraphics[height=4.5cm]{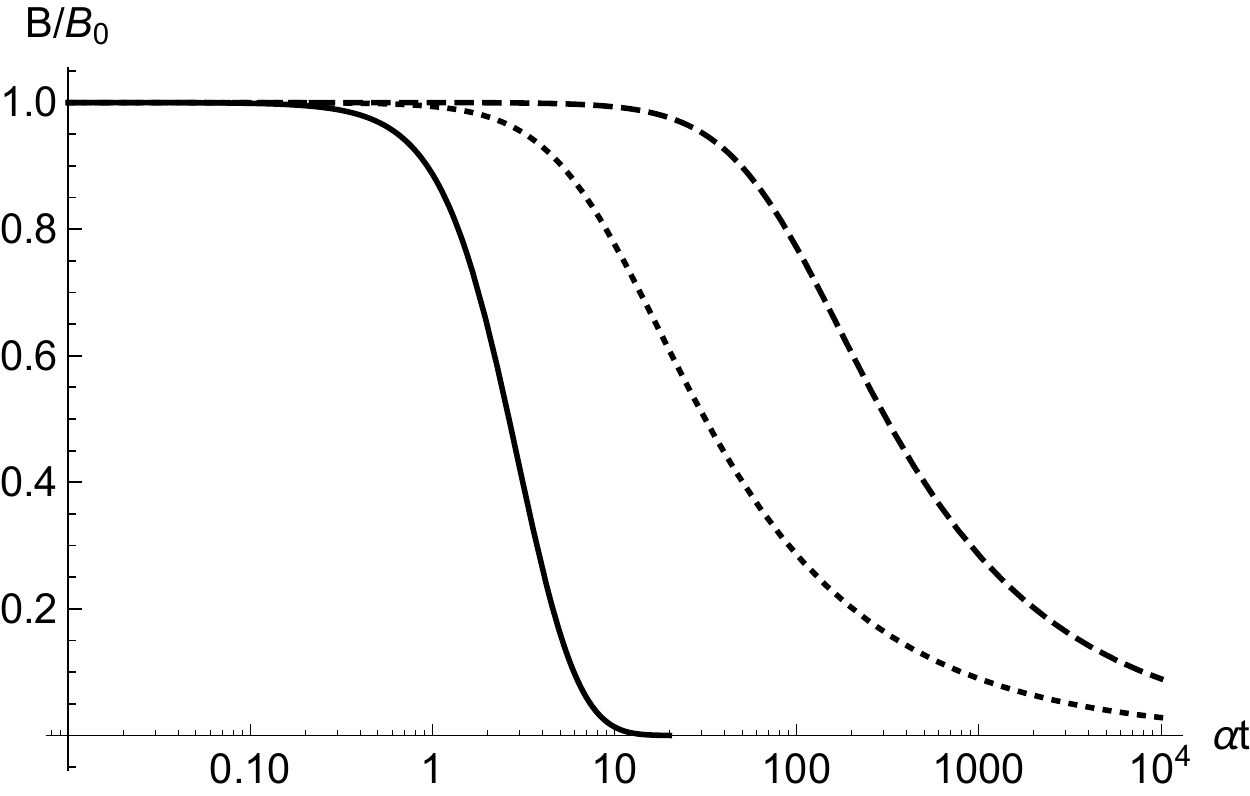} & &
      \includegraphics[height=4.5cm]{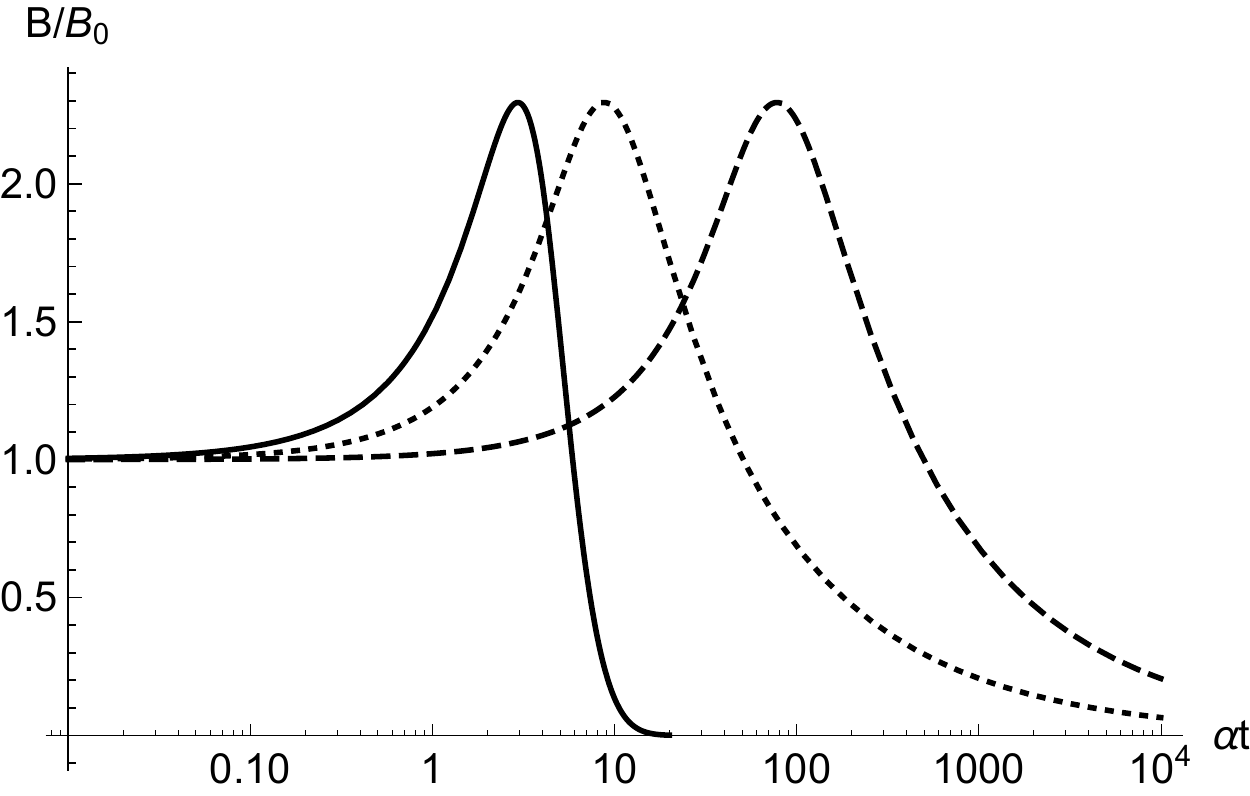}
      \end{tabular}
  \caption{Evolution of the magnetic field in hot medium for $ \sigma'=0$ (solid line),   $1$ (dotted line), $10$ (dashed line). The initial condition is $ \sigma'_\chi(0)\equiv \sigma'_0=0.5$, i.e.\ 50\% of the initial helicity is in the field (left panel) and $ \sigma'_0=0.95$, i.e.\ 95\% of the initial helicity is in medium, (right panel).}
\label{fig:b-hot}
\end{figure}

Unlike the vector potential, the magnetic field is not necessarily a monotonic function. Its time-dependence follows from \eq{b17} and is proportional to a product of an increasing and decreasing functions
\ball{g35}
B= k_0 A\propto \sqrt{\sigma_\chi(t)} a_0(t)\,.
\gal
The time $t_*$ at which $B(t)$ attains its maximum satisfies $\dot B(t_*)=0$ implying
\ball{g37}
\left[\dot\gamma +\frac{\dot\sigma'_\chi}{\sigma'_\chi}\right]_{\tau=\tau_*}=0\,.
\gal
Using \eq{b29} and \eq{g27} one derives 
\ball{g39}
\sigma'_\chi(\tau_*)= \frac{1}{2}\,.
\gal
Since $\sigma'_\chi(\tau)$ is monotonically decreasing from its initial value $\sigma'_0$, the magnetic field has maximum only if $\sigma_0'>1/2$, i.e.\ if most of the initial helicity is in the medium. Otherwise, $B(t)$ is a monotonically decreasing function of time (despite the fact that $A$ always grows). This is shown in \fig{fig:b-hot}.

The self-consistency of the adiabatic approximation employed in this section is verified in Appendix.

 \subsection{Insulating medium $\sigma=0$}

The chiral conductivity can be explicitly expressed as a function of $\tau$ in the case of vanishing electrical conductivity $\sigma=0$.  In this case,  solution to Eq.~\eq{g27} reads
\ball{k0}
\sigma'_\chi (\tau)= \frac{1}{1+\left(\frac{1}{ \sigma'_0}-1\right)e^{\tau }}\,.
 \gal
Clearly, at $\tau\ll 1$, the chiral conductivity is constant, while at $t>1$ it exponentially decreases with time, see \fig{fig:sigma-hot}. To compute the corresponding magnetic field  substitute \eq{k0} into \eq{b29}, which  yields 
\ball{k1}
\gamma(t)= \int_0^t\sigma_\chi(t')dt' = \tau - \ln\left[ e^{\tau}\left( 1-\sigma'_0\right)+\sigma'_0\right]\,.
\gal
Plugging this into \eq{b28} one derives for an insulating medium
\ball{k3}
a_0(t)= \frac{e^{\tau/2}}{\sqrt{e^{\tau}\left( 1-\sigma'_0\right)+\sigma'_0}}\,.
\gal
Since $\dot a_0>0$ for any $\sigma'_0>0$, magnetic  helicity $\mathcal{H}_\text{em}=a_0^2$  increases from its initial value $\mathcal{H}_\text{em}(0)=1$ to the final value of $\mathcal{H}_\text{em}(\infty)=(1-\sigma'_0)^{-1}=\mathcal{H}_\text{tot}$, where \eq{g23} has been used. 

In view of \eq{g31}, time-dependence of the vector potential  is given by
\ball{k5}
A\propto e^{\tau/2}\,.
\gal
It is an exponentially growing function of time as one would expect for an unstable state. See the solid line in right panel of  \fig{fig:sigma-hot}. 
Time-dependence of the corresponding magnetic field  follows from see \eq{b17}
\ball{k7}
B\propto
\frac{ e^{\tau/2}}{e^{\tau}(1-\sigma'_0)+\sigma'_0}\,.
\gal
At later times magnetic field exponentially decays even though vector potential grows exponentially. At $ \sigma'_0\le 1/2$ the magnetic field is a monotonically decreasing function of time. This is seen in the left panel of \fig{fig:b-hot}. It has a maximum only if $ \sigma'_0>1/2$, i.e.\ when most of the initial helicity is in the medium, in agreement with the general conclusion derived after \eq{g39}. In this case the maximum of magnetic field 
\ball{k9}
B(t_*)\propto \frac{1}{2\sqrt{ \sigma'_0(1- \sigma'_0)}}
\gal
occurs at $\tau_*= \ln[\sigma'_0/(1-\sigma'_0)]$. This is seen in the right panel of \fig{fig:b-hot}.

As have been already mentioned, for a finite total helicity  $\sigma_0'< 1$, therefore the right-hand-side of \eq{g27} does not vanish and  $\sigma_\chi$ is never a constant. Thus the divergence of the type shown in \eq{i13} never arises. It is always tamed by the helicity conservation.

\section{Magnetic field evolution in cold medium}\label{sec:n}

\begin{figure}[ht]
      \includegraphics[height=4.5cm]{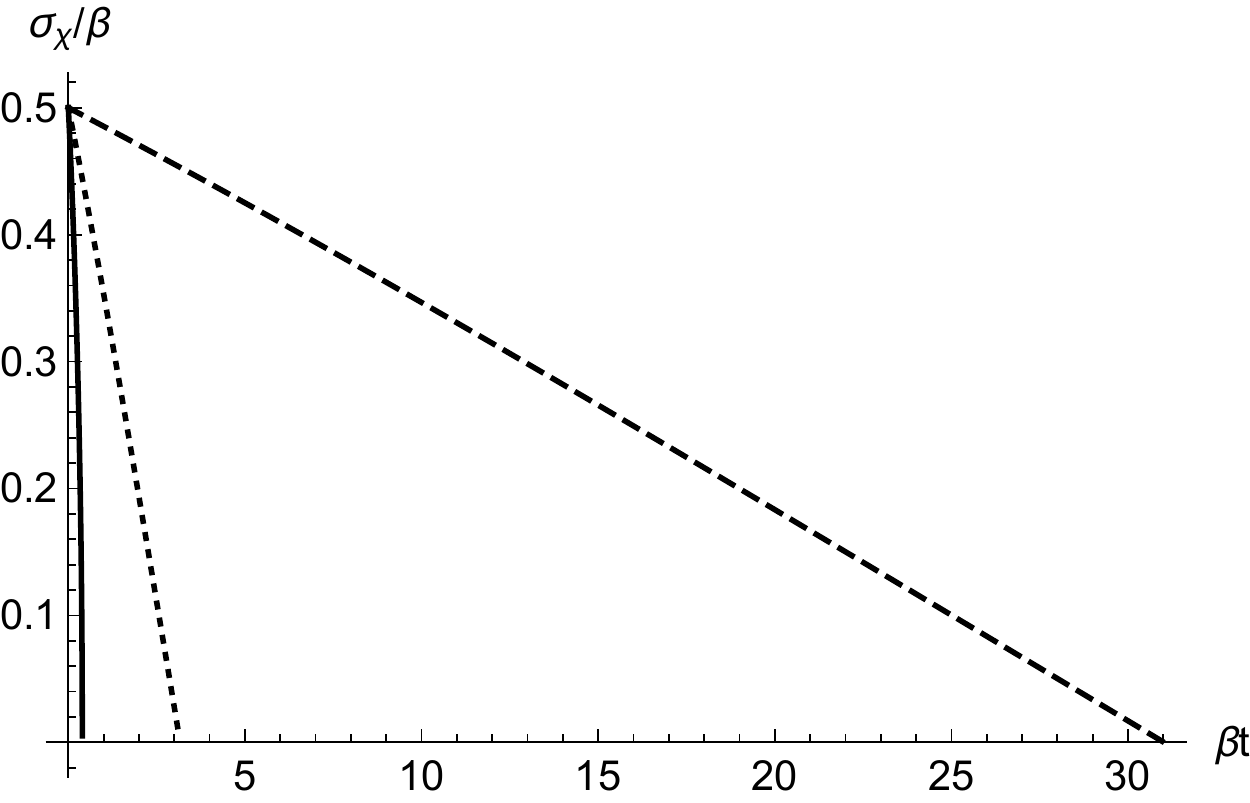} 
  \caption{Chiral conductivity of cold medium as a function of time for $\sigma_0'=0.5$. Lines represent solution of \eq{n10} at $\sigma'=0$ (solid line), $1$ (dotted line), $10$ (dashed line). }
\label{fig:sigma-chi-cold}
\end{figure}

In the cold medium, namely $T,\mu\ll \mu_5$,  the equation of state  is \cite{Fukushima:2008xe}
\ball{n1}
\mu_5^3 = 3\pi^2\aver{n_A}\,.
\gal
Together with the definition \eq{a22}, this implies that 
\ball{n3}
\aver{n_A}= \frac{ \sigma_\chi^3}{3\pi^2 c_A^3}\,.
\gal
Using this in \eq{g11} yields
\ball{n5}
\frac{\sigma_\chi^3(t)}{\beta^3}= 1-\frac{\mathcal{H}_\text{em}(t)}{\mathcal{H}_\text{tot}}\,,
\gal
where $\beta^3= 3\pi^2 c_A^4\mathcal{H}_\text{tot}/(2V)$ is a characteristic momentum scale. This time, in place of \eq{g23} one gets
\ball{n7}
\sigma_0= \beta(1-\mathcal{H}_\text{tot}^{-1})^{1/3}<\beta\,.
\gal
Once again using \eq{d18} in \eq{n5} one derives an equation obeyed by the chiral conductivity
\ball{n9}
3\sigma_\chi^2\, \dot \sigma_\chi= -\left(\sqrt{\sigma^2+\sigma^2_\chi}-\sigma\right) (\beta^3-\sigma_\chi^3)\,. 
\gal
It is convenient to define dimensionless quantities $\sigma'=\sigma_\chi/\beta$, $\sigma'=\sigma/\beta$ and $\tau = \beta t$ and write \eq{n9} as 
\ball{n10}
3\sigma_\chi'^2\, \dot \sigma_\chi'= -\left(\sqrt{\sigma'^2+\sigma'^2_\chi}-\sigma'\right) (1-\sigma_\chi'^3)\,,
\gal
where the dot means the $\tau$-derivative. As in the previous section, solution to this equation is bulky and is not recorded here. Rather it is shown in \fig{fig:sigma-chi-cold}. As in the hot medium, $\sigma_\chi$ is a decreasing function of time, which is also clear from \eq{n10} since its right-hand-side is negative. However, in contrast to \fig{fig:sigma-hot}, it vanishes not asymptotically at $\tau \to \infty$, but at a finial time $\tau_m$. 

\begin{figure}[ht]
\begin{tabular}{lcr}
      \includegraphics[height=4.5cm]{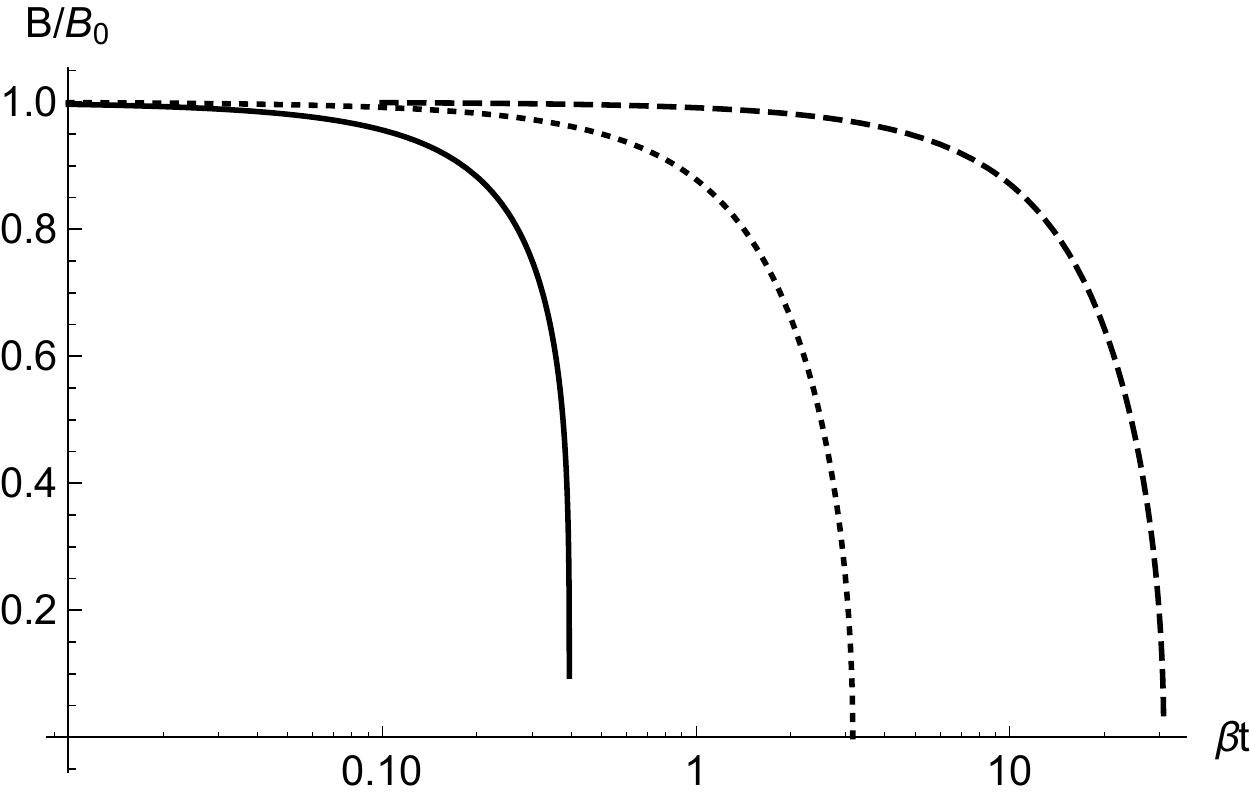} &&
      \includegraphics[height=4.5cm]{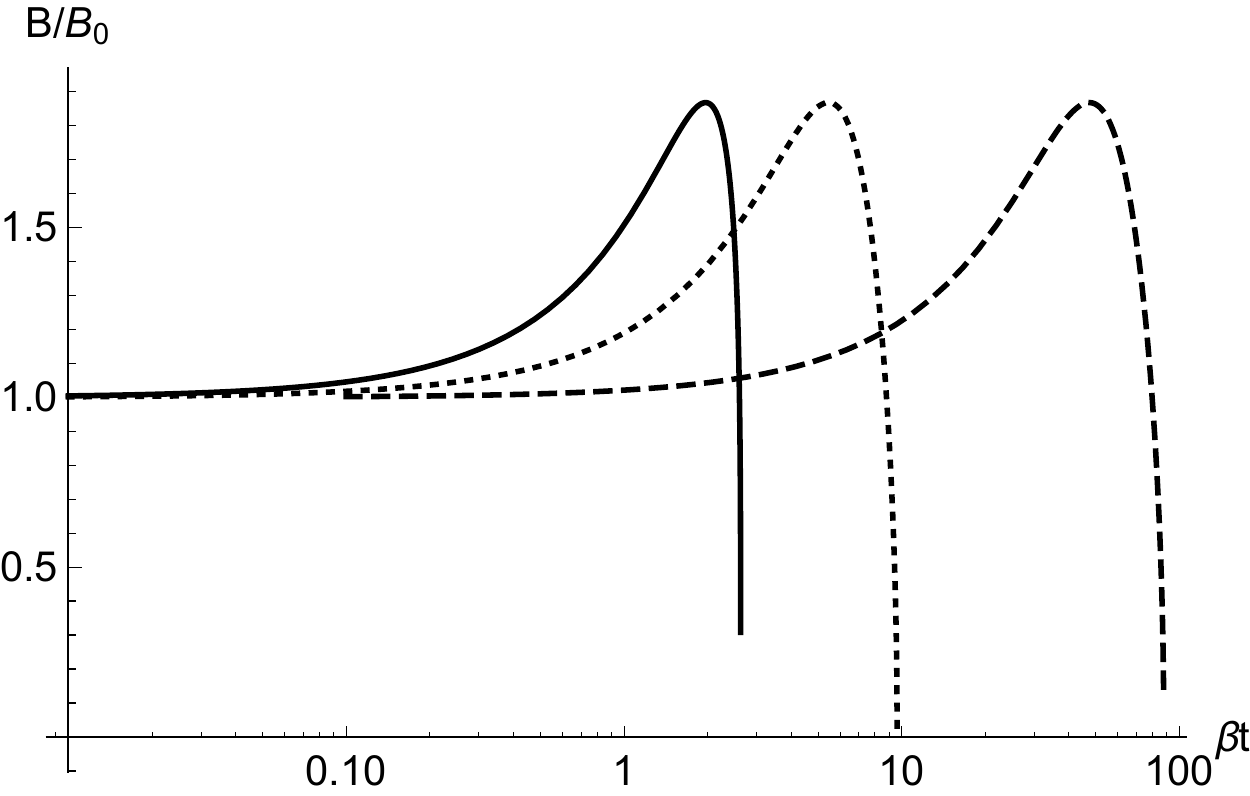}
      \end{tabular}
  \caption{Evolution of the magnetic field in cold medium for $ \sigma'=0$ (solid line),   $1$ (dotted line), $10$ (dashed line). The initial condition is $ \sigma'_\chi(0)\equiv \sigma'_0=0.5$, i.e.\ 50\% of the initial helicity is in the field (left panel) and $ \sigma'_0=0.95$, i.e.\ 95\% of the initial helicity is in medium, (right panel).}
\label{fig:b-cold-2}
\end{figure}

Time dependence of the vector potential and magnetic field can be computed employing \eq{g31}, \eq{g35}. They imply that the vector potential diverges  whereas the magnetic field vanishes at $\tau=\tau_m$. The results for the magnetic field are shown in \fig{fig:b-cold-2}. In order that $B(t)$ has a maximum, condition \eq{g37} must be satisfied. In the cold medium it yields
\ball{n13} 
\sigma'_\chi(t_*)= \frac{1}{2^{2/3}}\,.
\gal
Thus, if $\sigma_0'>1/2^{2/3}$, magnetic field has a maximum, whereas if $\sigma_0'<1/2^{2/3}$, it monotonically decreases. Thus, in order for the magnetic field to grow in the cold medium, more than 62\% of the initial helicity must be stored in  medium.

\subsection{Insulating medium $\sigma=0$}

In the insulating medium, evolution of the chiral conductivity is determined by the equation
\ball{n13}
3\sigma_\chi' \dot \sigma_\chi'= -(1-\sigma'^3_\chi)\,.
\gal
Its solution in a from of a transcendental equation for $\sigma_\chi'$ reads
\ball{n15}
\tau = \frac{1}{2}\ln \frac{(1-\sigma'_\chi)^2}{1+\sigma'_\chi+\sigma'^2_\chi }+\sqrt{3}\arctan\frac{2\sigma_\chi'+1}{\sqrt{3}}-\frac{1}{2}\ln \frac{(1-\sigma_0')^2}{1+\sigma_0'+\sigma'^2_0 }-\sqrt{3}\arctan\frac{2\sigma_0'+1}{\sqrt{3}}\,.
\gal
Explicit dependence of the  chiral conductivity $\sigma_\chi'$ on time  is shown in \fig{fig:sigma-chi-cold} by the solid line. It vanishes at $\tau_m$ that follows directly from \eq{n15} 
\ball{n17}
\tau_m = \frac{\pi}{6\sqrt{3}}-\frac{1}{2}\ln \frac{(1-\sigma_0')^2}{1+\sigma_0'+\sigma'^2_0 }-\sqrt{3}\arctan\frac{2\sigma_0'+1}{\sqrt{3}}\,.
\gal
At later times $\tau>\tau_m$, the chiral conductivity is identically zero implying that the chiral anomaly ceases to play any role in the magnetic field evolution. This is in contrast to the hot medium, where  $\sigma_\chi'$ vanishes only asymptotically at $\tau\to \infty$. 

\subsection{Conducting medium $\sigma\gg \sigma_\chi$}

In this limit \eq{n10} reduces to 
\ball{n33}
\dot \sigma_\chi'= -\frac{1}{6\sigma'}(1-\sigma'^3_\chi)\,,
\gal
and is solved by
\ball{n35}
\frac{\tau}{\sigma'}=\ln \frac{(1-\sigma_\chi')^2}{1+\sigma_\chi'+\sigma'^2_\chi} -\frac{6}{\sqrt{3}}\arctan\frac{\sqrt{3}\sigma_\chi'}{2+\sigma_\chi'}- \ln \frac{(1-\sigma'_0)^2}{1+\sigma'_0+\sigma'^2_0} +\frac{6}{\sqrt{3}}\arctan\frac{\sqrt{3}\sigma_0'}{2+\sigma_0'}\,.
\gal
The evolution of the chiral conductivity proceeds from $\tau=0$ up to $\tau =\tau_m$ where $\sigma_\chi'$ vanishes. Thus $\tau_m/\sigma'$ is given by the last two terms in \eq{n35}. Notice that $\tau_m$ is proportional to the medium conductivity. This is seen in \fig{fig:sigma-chi-cold}.

\section{Discussion and summary}\label{sec:s}

Development of magnetic field instability in a chiral medium has been the main subject of this paper. For a qualitative analytical understanding of the instability dynamics, we developed a model that approximates the vector potential by the fastest growing helicity state. Maxwell equations with anomalous currents determine dependence of the magnetic field on the chiral conductivity. They however, do not determine the functional form of $\sigma_\chi(t)$ and neither does the energy conservation requirement, as was verified in \sec{sec:d}. It is the chiral anomaly equation \eq{g3} and the resulting helicity conservation \eq{g11}, in conjunction with the equation of state that fix the time-dependence of the chiral conductivity. 

Two equations of state were considered \eq{g17} and \eq{n1}, which are referred to as the hot and cold matter respectively. Whether the magnetic field increases at the early stages of its evolution depends on a single parameter, the fraction of the total helicity initially in the medium $\sigma_0'$, which  also happens to be the initial value of the chiral conductivity (in dimensionless units). In hot medium it must be more than 1/2 while in the cold medium more than $1/2^{2/3}$. This parameter also determines the instant $t_*$ when the maximum occurs and the peak strength of  the magnetic field $B(t_*)$. If $\sigma_0'$ is below the values indicated above, the magnetic field is a monotonically decreasing function of time. This is seen in \fig{fig:b-hot} and \fig{fig:b-cold-2}. Magnetic helicity always monotonically increases from unity at the initial time to $\mathcal{H}_\text{tot}$ at the later times. This is because in the present model helicity flows only from the medium to the magnetic field.  The eventual vanishing of the chiral conductivity at late times and the transfer of all helicity to magnetic field is a general property of the MCS theory in homogeneous medium, even though the details of the time-evolution are model-dependent. This is because in a stationary state of the chiral medium with a finite value of  $\sigma_\chi=\sigma_\infty$, the magnetic field  contains unstable modes with $k<\sigma_\infty$, which increase the magnetic helicity, thereby driving $\sigma_\chi$ to smaller values due to the helicity conservation \eq{g9}. The stability is achieved only when $\sigma_\infty=0$.\footnote{This conclusion may change in the presence of the helicity non-conserving processes such as discussed in \cite{Tuchin:2016qww}.}

Duration of the chiral evolution strongly depends on the equation of state. In hot matter, which is appropriate for the description of the quark-gluon plasma produced in relativistic heavy-ion collisions, the magnetic helicity approaches the total helicity only asymptotically at $t\to \infty$. However, in cold matter,  the chiral conductivity vanishes at finite time $\tau_m$ that depends on  $\sigma_0'$ and the electrical conductivity. At this time, the magnetic field vanishes, while the vector potential diverges (this is why the magnetic helicity is finite).  

Our results for the hot matter are qualitatively similar to the numerical results of \cite{Hirono:2015rla} (stages II and III). One however, has to bear in mind that \cite{Hirono:2015rla} neglected the second time derivative of $A$ in \eq{a24}, which is important at not very large electrical conductivity. Another numerical calculation of the time-evolution of magnetic helicity in chiral medium was reported in \cite{Manuel:2015zpa}. There a different approximation was adopted which precludes the direct comparison with the results of the present work.

\acknowledgments
The author wishes to  thank Dmitri Kharzeev  for helpful communications. This work  was supported in part by the U.S. Department of Energy under Grant No.\ DE-FG02-87ER40371.

\appendix
\section{Verification of the adiabatic approximation}\label{appA}

\begin{figure}[ht]
      \includegraphics[height=4.5cm]{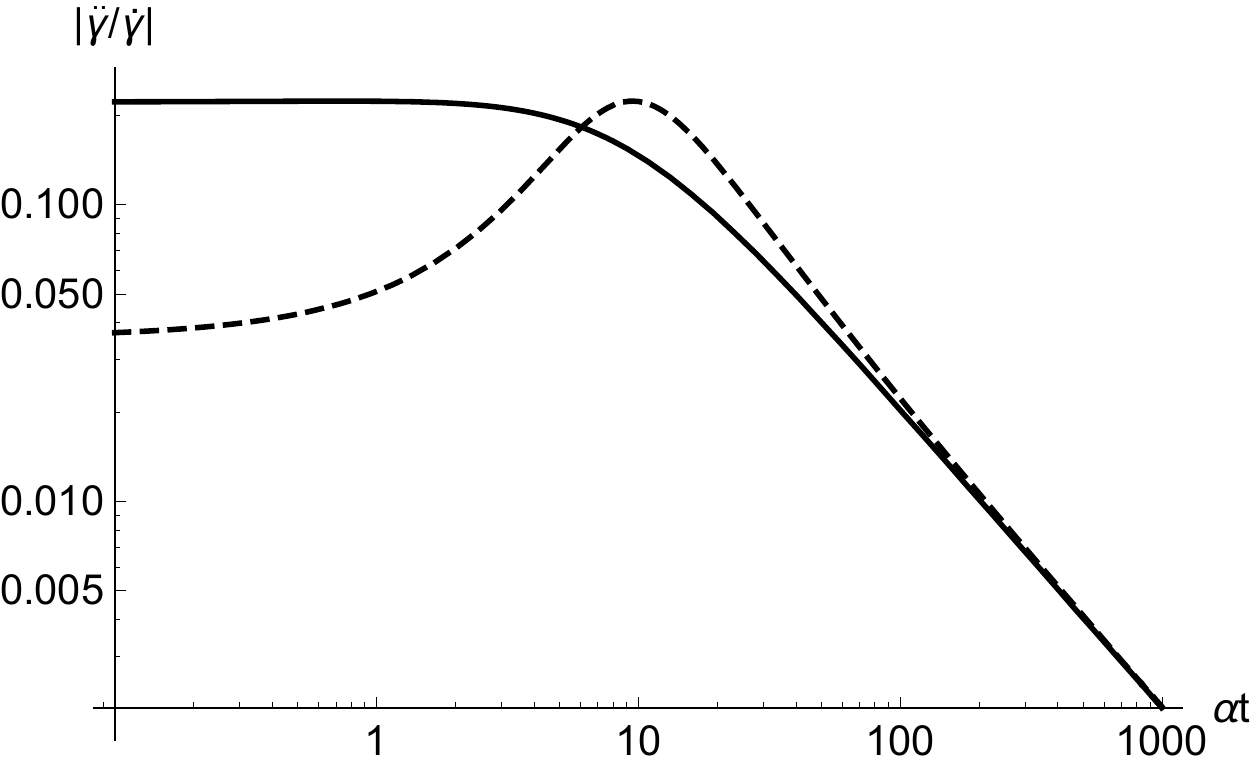} 
  \caption{Accuracy of the adiabatic approximation for $\sigma'=1$ and  $\sigma'_0=0.5$ (solid line) and  $0.95$ (dashed line) in hot medium. }
\label{fig:wkb}
\end{figure}

The adiabatic approximation adopted for the analysis of the time evolution in this paper assumes that $\ddot \gamma\ll \dot \gamma$, see the text following Eqs.~\eq{b22} and \eq{d18}. Plotted in \fig{fig:wkb} is the ratio $|\ddot\gamma/\dot \gamma|$ for certain values of conductivity. Similar results hold for other values of the parameters. Clearly, the adiabatic approximation is self-consistent.


\end{document}